\documentclass[lettersize,journal]{IEEEtran}
\usepackage{amsmath,amsfonts}
\usepackage[caption=false,font=normalsize,labelfont=sf,textfont=sf]{subfig}
\usepackage{textcomp}
\usepackage{stfloats}
\usepackage{url}
\usepackage{verbatim}
\usepackage{graphicx}
\usepackage{cite}

\usepackage{xcolor}
\usepackage{amsmath,amssymb, amsfonts, mathtools}
\mathtoolsset{showonlyrefs}
\usepackage[utf8]{inputenc}
\usepackage[english]{babel}
\usepackage{mathrsfs}
\usepackage{algorithm}
\usepackage[noend]{algpseudocode}
\usepackage{hyperref}

\usepackage{amsmath,amsthm,amssymb}
\usepackage{mathtools}
\mathtoolsset{showonlyrefs}
\usepackage{mathrsfs}
\usepackage{comment}
\newtheorem{theorem}{Theorem}

\DeclareMathOperator{\co}{co}
\DeclareMathOperator*{\argmin}{arg\,min}
\DeclareMathOperator*{\argmax}{arg\,max}
\newcommand{\norm}[1]{\left\lVert#1\right\rVert}

\newcommand{\mc}{\mathcal}

\definecolor{blue}{rgb}{0.0, 0.0, 1.0}

\begin{document}

\title{Safe Reinforcement Learning using Robust Control Barrier Functions}

\author{Yousef Emam,$~\IEEEmembership{Student Member,~IEEE,}$ Gennaro Notomista,$~\IEEEmembership{Member,~IEEE,}$ Paul Glotfelter,$~\IEEEmembership{Member,~IEEE,}$\\ Zsolt Kira,$~\IEEEmembership{Senior Member,~IEEE,}$ and Magnus Egerstedt,$~\IEEEmembership{Fellow,~IEEE}$
	\thanks{This work was supported by the Army Research Lab through ARL DCIST CRA W911NF-17-2-0181.}
	\thanks{Y. Emam, Z. Kira are with the Institute for Robotics and Intelligent Machines, Georgia Institute of Technology, Atlanta, GA 30332, USA {\tt\small emamy@gatech.edu, zsolt.kira@gtri.gatech.edu}}
	\thanks{G. Notomista is with the Department of Electrical and Computer Engineering, University of Waterloo, Waterloo, ON N2L 3G1, Canada {\tt\small gennaro.notomista@uwaterloo.ca}}
	\thanks{P. Glotfelter is with Outrider, 16163 W. 45th Drive, Golden, CO 80403, USA {\tt\small pglotfel@gmail.com}}
	\thanks{M. Egerstedt is with the Samueli School of Engineering, University of California, Irvine, CA 92697, USA {\tt\small magnus@uci.edu}}
}



\maketitle

\begin{abstract}
    Reinforcement Learning (RL) has been shown to be effective in many scenarios. However, it typically requires the exploration of a sufficiently large number of state-action pairs, some of which may be unsafe. Consequently, its application to safety-critical systems remains a challenge. An increasingly common approach to address safety involves the addition of a safety layer that projects the RL actions onto a safe set of actions. In turn, a difficulty for such frameworks is how to effectively couple RL with the safety layer to improve the learning performance. In this paper, we frame safety as a differentiable robust-control-barrier-function layer in a model-based RL framework. Moreover, we also propose an approach to modularly learn the underlying reward-driven task, independent of safety constraints. We demonstrate that this approach both ensures safety and effectively guides exploration during training in a range of experiments, including zero-shot transfer when the reward is learned in a modular way. 
\end{abstract}

\begin{IEEEkeywords}
Robust/Adaptive Control, Reinforcement Learning, Robot Safety.
\end{IEEEkeywords}

\section{Introduction}


\IEEEPARstart{R}{einforcement} Learning (RL) has been used successfully in a number of robotics applications \cite{kober2013reinforcement, arulkumaran2017deep}. Typically, an RL agent must explore a sufficiently large number of states during training to learn a meaningful policy. However, this exploration may involve visiting unsafe states, making RL unsuitable for \textit{safety-critical} systems where the cost of failure is high \cite{9392290, cowen2020samba}.  As such, \textit{safe reinforcement learning} is a growing field of research aimed at tackling this challenge \cite{garcia2015comprehensive} by combining control-theoretic techniques with RL. A large variety of safe-RL approaches have been proposed in the literature. These approaches can be categorized as either approaches that solely focus on obtaining a safe policy \textit{at the end} of training (e.g., \cite{altman1998constrained, geibel2005risk, chow2017risk}) and approaches, including the work in this paper, that focus on safe exploration \textit{during} training (e.g., \cite{cheng2019end, berkenkamp2017safe, chow2018lyapunov}). 

Safe training for RL inherently consists of two components: a constraint-based component, which ensures the agent's safety, and a performance metric, typically captured by a reward or cost function, that the agent aims to optimize. In the work from \cite{chow2018lyapunov, chow2019lyapunov}, the authors propose a Lyapunov-based approach for solving Constrained MDPs (CMDPs) \cite{altman1999constrained} which guarantees the safety of the agent during training. Another similar approach is the work in \cite{berkenkamp2017safe} where the authors provide a model-based RL framework for safely expanding a region of attraction for the agent. This region of attraction is a set of states for which a stabilizing policy is known. However, the main drawback of these approaches is that they tend to be quickly intractable for more complex environments. Specifically, the method in \cite{berkenkamp2017safe} relies on a discretization of the state space to check states at different level sets of the Lyapunov function, an approach which cannot scale to high dimensional states. Along similar lines, the approach in \cite{chow2018lyapunov} relies on an optimization at each step  with as many constraints as there are states, each of which involves an integral over the entire action space. 

Another category of approaches leverages a model-based safety framework which serves to prevent the exploration of unsafe states (e.g., \cite{fisac2018general, li2018safe}) by projecting the action taken by the RL agent onto a safe-set of actions. In turn, a central challenge to these approaches is how to leverage the knowledge of the dynamics and the model-based safety component to effectively guide exploration during training. Towards this end, one such approach is that in \cite{cheng2019end}, where the authors assume knowledge of a Control Barrier Function (CBF) along with a prior on the dynamics model. During training, the uncertainty in the dynamics is learned using Gaussian Processes (GPs) and accounted for in the CBF-based safety layer. Moreover, a novelty in that work is the addition of a supervised-learning component to the policy  referred to by the authors as the compensator, which is a neural net that is used to learn the output of the CBF.

However, the approach in \cite{cheng2019end} suffers from three drawbacks. First, the discrete-time CBF formulation used is restrictive in that it is only amenable to real-time control synthesis via quadratic programming for affine CBFs (i.e., CBFs of the form $h(x)=p^Tx+q$ where $x$ is the state). For example, in the context of collision avoidance, affine CBFs can only encode polytopic obstacles. Moreover, the supervised learning of the CBF layer fundamentally results in an approximation, which, in the worst case, could negatively affect the learning of the RL agent if inaccurate. Lastly, the modifications to the actions made by the CBF layer and supervised learning components are not explicitly accounted for in the loss functions of the RL algorithm. This is equivalent to the effects of the CBF layer being treated as part of the unknown transition function with respect to the RL algorithm and can thus impede learning in cases where the agent behaves aggressively to avoid safety violations.

To address these challenges, by leveraging recent advances in differentiable optimization \cite{amos2017optnet, agrawal2019differentiable}, we introduce a differentiable Robust CBF (RCBF) \cite{emam2019robust} based safety layer that is compatible with standard policy-gradient RL algorithms. RCBFs are amenable to real-time control synthesis, even if the functions are non-affine, and can encode a wide class of disturbances on the dynamics making them applicable to a large variety of systems. Additionally, the differentiable safety layer eliminates the need to learn the behavior of the CBF and thus a supervised-learning component. 

Moreover, it may often be the case that the CBF constraints during training and testing differ when dealing with safety critical systems. Such cases may include a drone that must remain within a given distance from a safety operator in applications such as construction \cite{li2019applications} or entertainment \cite{kim2018survey}. Moreover, training and testing constraints also differ when treating RL training as a persistent task which involves the addition of charging constraints \cite{notomista2020persistification}, or when using multiple agents to speed up training (e.g., \cite{makoviychuk2021isaac}) in which case inter-agents collisions must be ensured. In such scenarios, the learned value function at the end of the training needs to be independent of a subset of the CBF constraints imposed during training. Towards this end, we propose an approach for learning the reward-driven task modularly which is empirically validated in Section~\ref{sec:experiments}. Specifically, we test the zero-transfer ability of various approaches in environments requiring different safety constraints and demonstrate that the proposed approach significantly improves performance.   

\section{Background Material} 
\label{sec:background}

In this section, we introduce CBFs and their robust counterpart from \cite{emam2019robust} which can be embedded as a wrapper around any nominal controller to ensure the safety of a disturbed dynamical system. Then, we discuss the underlying RL algorithm, namely Soft Actor Critic (SAC) \cite{haarnoja2018soft}, that we utilize as the nominal controller in our framework. The motivation behind using SAC is that it is sample-efficient; however, we note that the methods proposed in this paper are compatible with any off-the-shelf policy-gradient algorithm.


We consider the following disturbed control-affine system
\begin{equation}
\label{eq:main_system}
    \dot{x}(t) = f(x(t)) + g(x(t)) u(x(t)) + d(x(t)),
\end{equation}
where $f : \mathbb{R}^{n} \to \mathbb{R}^{n}$ denotes the drift dynamics, $g : \mathbb{R}^{n} \to \mathbb{R}^{n \times m}$ is the input dynamics, $d : \mathbb{R}^{n} \to \mathbb{R}^{n}$ is an unknown deterministic disturbance and $u : \mathbb{R}^{n} \to \mathbb{R}^{m}$ is the input signal. We assume that $f$, $g$, $d$ and $u$ are continuous. We note that the work in this paper can be straightforwardly extended to the case where $d$ is stochastic since GPs are used to learn the disturbance.

\subsection{Robust Control Barrier Functions}
\label{subsec:rcbfs}

Control barrier functions \cite{ames2014,xu2015,AmesBarriers,ogren2006autonomous} are formulated with respect to control-affine systems
\begin{equation}
    \label{eq:control-affine}
    \dot{x}(t) = f(x(t)) + g(x(t))u(x(t)) , x(0) = x_{0}.
\end{equation}
A set $\mathcal{C}$ is called forward invariant with respect to \eqref{eq:control-affine} if given a (potentially non-unique) solution  to \eqref{eq:control-affine} $x : [0, t_{1}] \to \mathbb{R}^{n}$,
$x_{0} \in \mathcal{C} \implies x(t) \in \mathcal{C}, \forall t \in [0, t_{1}]$. Note that \eqref{eq:control-affine} is the unperturbed version of \eqref{eq:main_system}.



Since we are interested in guaranteeing the safety of the disturbed dynamical system \eqref{eq:main_system},  we leverage RCBFs \cite{emam2019robust, emam2021data},  which generalize the notion of CBFs to systems obeying the following differential inclusion
\begin{equation} 
    \label{eq:control-affine-disturbed}
    \dot{x}(t) \in f(x(t)) + g(x(t))u(x(t)) + D(x(t)), x(0) = x_{0} ,
\end{equation}
where $f$, $g$, $u$ are as in \eqref{eq:main_system} and $D : \mathbb{R}^{n} \to 2^{\mathbb{R}^{n}}$ (the disturbance) is an upper semi-continuous set-valued map that takes nonempty, convex, and compact values. Note that $2^{\mathbb{R}^{n}}$ refers to the power set of $\mathbb{R}^{n}$ and that the assumptions made on the disturbance are conditions to guarantee the existence of solutions \cite{cortes2008discontinuous}.

Moreover, it was shown that for a specific form of $D$, we can recover a similar formulation of regular CBFs 
 with almost no additional computational cost as stated in the following theorem.  An important aspect of this theorem is that forward invariance is guaranteed for all trajectories of \eqref{eq:control-affine-disturbed}.
\begin{theorem}{\cite{emam2019robust}}
    \label{thm:rcbf-add}
    Let $h : \mathbb{R}^{n} \to \mathbb{R}$ be a continuously differentiable function.  Let $\psi_{i} : \mathbb{R}^{n} \to \mathbb{R}^{n}$, $i \in \{1, \hdots, p\}$ be a set of $p > 0$ continuous functions, and define the disturbance $D : \mathbb{R}^{n} \to 2^{\mathbb{R}^{n}}$ as 
    \begin{equation}
        D(x') = \co \Psi(x') = \co\{\psi_1(x')\ldots\psi_p(x')\}, \forall x' \in \mathbb{R}^{n},
    \end{equation}
    where $\co$ denotes the convex hull. If there exists a continuous function $u : \mathbb{R}^{n} \to \mathbb{R}^{m}$ and a locally Lipschitz extended class-$\mathcal{K}$ function $\gamma : \mathbb{R} \to \mathbb{R}$ such that 
    \begin{equation} 
        \label{eq:mainTheorem}
            \begin{split}
                L_{f}h(x') + L_{g}h(x')u(x') \geq 
                -\gamma(h(x')) \\ \qquad - \min \nabla h(x')^{\top} \Psi(x'), \forall x' \in \mathbb{R}^{n} ,
            \end{split}
    \end{equation}
    then $h$ is a valid RCBF for \eqref{eq:control-affine-disturbed}.
\end{theorem}

In other words, given the set $D$, RCBFs ensure safety of all trajectories of the disturbed system \eqref{eq:control-affine-disturbed}. Moreover, the result in Theorem~\ref{thm:rcbf-add} can be straightforwardly extended to the case where $D$ is a finite union of convex hulls as discussed in \cite{emam2019robust}, which can be leveraged to capture non-convex disturbances. As such, we can learn an estimate of the unknown disturbance $d$ from \eqref{eq:main_system} and leverage RCBFs to ensure the safety with respect to that estimate. 

\subsection{Soft Actor-Critic (SAC)} 
\label{subsec:sac}

The underlying RL problem considered is a policy search in an MDP defined by a tuple ($\mc X$, $\mc U$, $f$, $g$, $d$, $r$), where $\mc X~\subset~\mathbb{R}^n$ denotes the state space, $\mc U~\subset~\mathbb{R}^m$ is the input space, $f$, $g$, $d$ are as in \eqref{eq:main_system}, and $r \colon \mc X \times \mc U  \times \mc X \rightarrow \mathbb{R}$ is the reward associated with each transition. Note that the state transitions for the MDP are obtained by discretizing the dynamics \eqref{eq:main_system} as
\begin{equation}
\label{eq:discrete_dynamics}
    x_{k+1} = \Delta t (f(x_{k}) + g(x_{k})u(x_{k}) + d(x_k)) + x_k, 
\end{equation}
where $x_{k}$ denotes the state at time step $k$ and $\Delta t$ is the time step size. We note that this approximation has no effect on safety, which is ensured using RCBFs as discussed in Section~\ref{subsec:safety_layer}. 


Although the proposed approach can be straightforwardly extended to any off-the-shelf policy-gradient algorithm, in this paper, we chose to focus SAC and utilize it as the underlying RL algorithm since it is state-of-the-art in terms of sample efficiency \cite{haarnoja2018soft} and thus aligns with the objective of this work. SAC maximizes an entropy objective which is given by the following 
\begin{equation}
\pi^{\ast} = \argmax \sum_t
\mathbb{E}_{(x_t, u^{RL}_t) \sim \rho_{\pi}} [r(x_t, u^{RL}_t) + \alpha_{e} \mc H (\pi (\cdot | x_t))], 
\end{equation}
where $x_t$ and $u^{RL}_t$ denote the state and action sampled from the policy at timestep $t$, respectively, $\rho_{\pi}$ denotes the distribution of states and actions induced by the policy $\pi$, and the term $\mc H (\pi (\cdot | x_t))$ is the entropy term which incentivizes exploration and $ \alpha_{e} > 0$ is a weighting parameter between the two terms in the objective.

The algorithm relies on an actor-critic approach with a fitted Q-function parametrized by $\theta$ and a fitted actor $\pi$ parametrized by $\phi$. As such, the critic loss is given by
\begin{align}
\label{eq:q_update}
    J_{Q}(\theta) = \mathbb{E}_{(x_t, u^{RL}_t) \sim \mc{D}_R} &[ \frac{1}{2}  (Q_{\theta}(x_t, u_t) - (r(x_t, u^{RL}_t) \\
    &+ \gamma \mathbb{E}_{x_{t+1} \sim p} [V_{\bar{\theta}}(x_{t+1})]))^{2}],
\end{align}
where
\begin{equation}
V_{\bar{\theta}}(x_{t}) = \mathbb{E}_{ u^{RL}_t \sim \pi_{\phi}} [ Q_{\bar{\theta}}(x_t, u^{RL}_t) - \alpha_{e} \log{\pi_{\phi}(u^{RL}_t | x_t)}], 
\end{equation}
$\mc{D}_R$ is the replay buffer and $\bar{\theta}$ is the target Q-network parameters. Finally, the policy loss is given by
\begin{equation}
\label{eq:pi_update}
\begin{aligned}
J_{\pi}(\phi) &= \mathbb{E}_{ x_t \sim \mc{D}_R} [\mathbb{E}_{ u^{RL}_t \sim \pi_{\phi}} [\alpha_{e} \log{\pi_{\phi}(u^{RL}_t | x_t)} \\&\quad- Q_{\theta}(x_t, u^{RL}_t)]].
\end{aligned}
\end{equation}



\section{Main Approach}
\label{sec:main}

\begin{figure}
    \centering
    \includegraphics[width=.85\linewidth]{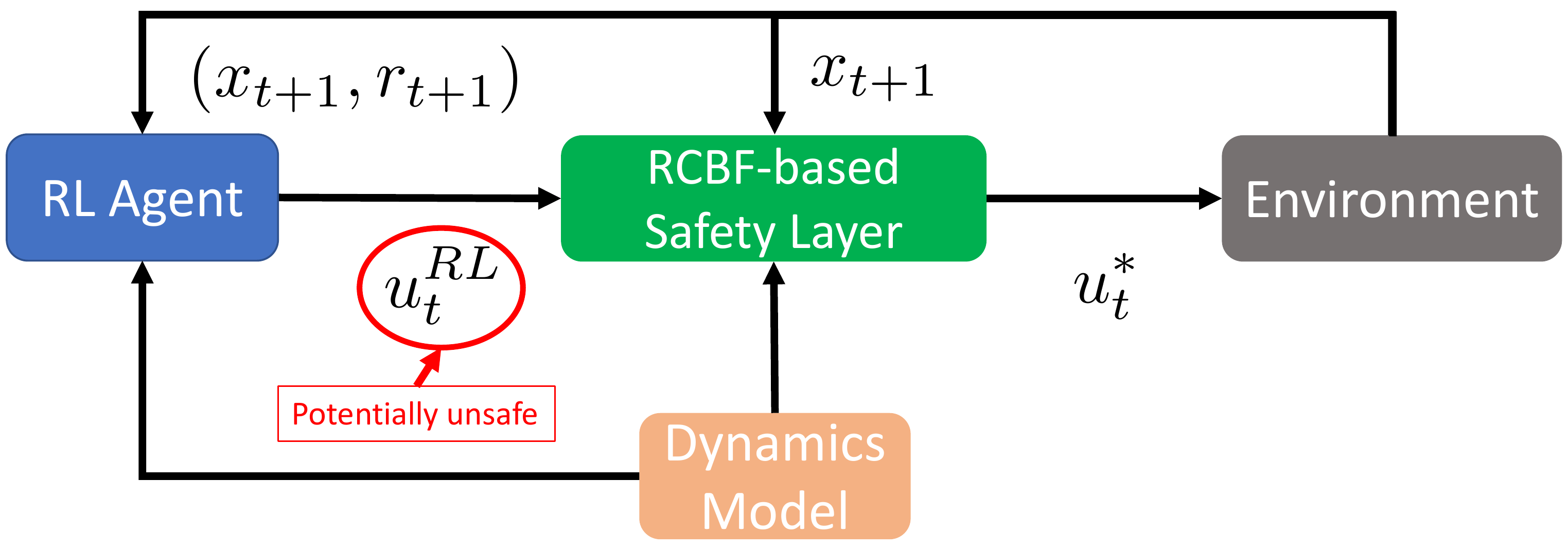}
    \caption{A diagram of the proposed framework. At time step $t$ the RL agent outputs the potentially unsafe control $u^{RL}_t$ which is rendered safe using the CBF controller. The safe control input $u^{\ast}_t$ given by \eqref{eq:final_action} is then applied in the environment. Note that the dynamics model is used by both the CBF controller to guarantee safety, and the RL agent to increase its learning's sample efficiency if possible.}
    \label{fig:framework-diagram}
\end{figure}

In this section, we introduce a novel framework which we refer to as SAC-RCBF and discuss methods to improve its learning performance. Specifically, the first subsection discusses how GPs can be leveraged to learn the disturbance $D(\cdot)$ from \eqref{eq:control-affine-disturbed}. Then, we introduce the RCBF-based optimization program which leverages the disturbance estimates to minimally alter the RL policy's actions to ensure safety of the system as pictorially depicted in Figure~\ref{fig:framework-diagram}. Subsections~\ref{subsec:sample-efficiency} and \ref{subsec:mod-task-learning} discuss methods to improve the sample efficiency and transfer ability of SAC-RCBF respectively.

\subsection{Disturbance Estimation} 
\label{subsec:disturb_est}

Although RCBFs are compatible with various data-driven methods, in this paper, we choose to focus on estimating the disturbance set $D(x)$ using GPs \cite{rasmussen2003gaussian}. This learning is achieved by obtaining a dataset $\mathcal{D}~=~\{x^{(i)}, y^{(i)}\}^{N}_{i=1}$ with labels $y^{(i)}$ given by
$
    y^{(i)} = \hat{\dot{x}}^{(i)} - f(x^{(i)}) - g(x^{(i)})u^{(i)},
$
where $\hat{\dot{x}}^{(i)}$ is the noisy measurement of the dynamics during exploration. Note that, in this case, $y^{(i)} \in \mathbb{R}^n$; therefore, we train one GP per dimension for a total of $n$ GP models.

In turn, we can obtain the disturbance estimate for a query point $x_{\ast}$ as 
\begin{equation}
\label{eq:disturb_estimate}
[{D}(x_{\ast})]_d = \mu_d(x_{\ast}) + [-k_c \sigma_d(x_{\ast}), k_c \sigma_d(x_{\ast})],
\end{equation}
where $\mu_d(x_{\ast})$ and $\sigma_d(x_{\ast})$ are the mean and standard-deviation predictions of the $d$-th GP for query point $x_{\ast}$. The coefficient $k_c$ is a user-chosen confidence parameter (e.g., $k_c~=~2$ achieves a confidence of $95.5\%$). Note that this is a different representation of $D(x)$ than the one used by Theorem~\ref{thm:rcbf-add} where the disturbance is the convex hull of a set of points. However, the desired form can be readily obtained by defining $\Psi$ as the $2^{n}$ vectors generated by permuting the entries of $D$ from \eqref{eq:disturb_estimate}. 

\subsection{RCBF-based Safety Layer}
\label{subsec:safety_layer}

In this subsection, we present the minimally invasive RCBF Quadratic Program (RCBF-QP) based control synthesis framework which guarantees the safety of the RL agent with dynamics as in \eqref{eq:control-affine-disturbed}. As depicted in Figure~\ref{fig:framework-diagram}, given the state, the disturbance estimate from the GPs and the action of the RL agent, this safety layer minimally alters the action so as to ensure safety. Specifically, the RCBF compensation term $u^{S}$ is given by
\begin{align}
    \label{eq:QP}
    & (u^{S}(x', u^{RL}(x')), \epsilon_s) = \argmin_{u, \epsilon} \|u\|^2 + l\epsilon_s^2\\ 
    & \text{s.t. } \nabla h_i(x')^{\top}(f(x') + g(x')(u(x') + u^{RL}(x'))) \geq \\ & \quad -\alpha(h_i(x')) 
     - \min \nabla h_i(x')^{\top}\Psi(x') + \epsilon_s, \quad \forall i \in \mc{N}_s,
\end{align}

\noindent
where $u^{RL}(x') \sim \pi_{\phi}(\cdot | x')$ is the output of the RL policy at state $x'$, $\epsilon_s \in \mathbb{R}$ is a slack-variable that serves to ensure feasibility of the QP, $l > 0$ is a large weighting term to minimize safety violations and $\mc{N}_s = \{1, \ldots, n_s\}$ where $n_s$ is the number of RCBFs. As depicted in Figure~\ref{fig:framework-diagram}, the final safe action taken in the environment is given by 
\begin{equation}
\label{eq:final_action}
    u^{*}(x') = u^{S}(x', u^{RL}(x')) + u^{RL}(x').
\end{equation}
Note that constraints can be modularly added to \eqref{eq:QP} to also account for actuator limits. Moreover, the need for a slack variable stems from cases where no control input can satisfy the control barrier certificate and a safety violation is inevitable. We refer the reader to \cite{ames2019control, wang2017safety, squires2018constructive} for approaches on how to construct CBFs that guarantee  the  existence  of  solutions  under  actuator constraints which typically leverage a back-up safety maneuver.

As discussed in Section~\ref{subsec:rcbfs}, RCBFs are formulated with respect to the continuous-time differential inclusion \eqref{eq:control-affine-disturbed}. However, as highlighted in Theorem~$3$ of \cite{ames2016control}, under certain Lipschitz-continuity assumptions, solving the RCBF-based optimization program at a sufficiently high frequency ensures safety of the system during inter-triggering intervals as well. As such, if $\Delta t$ from \eqref{eq:discrete_dynamics} is sufficiently small, we can query the RL policy and solve the RCBF-QP at the required rate for ensuring safety at all times. On the other hand, if $\Delta t$ is relatively large, safety can still be enforced by applying a zero-order hold on the RL action and subsequently solving the RCBF-QP at the necessary rate; however, in the interest of brevity, we leave these modifications to future work.

\subsection{Sample Efficiency}
\label{subsec:sample-efficiency}
As mentioned in the Introduction, naively coupling SAC with the RCBF-QP suffers from drawbacks. Specifically, the effects of the RCBF-QP are not accounted for in the RL policy's training and the partially-learned transition model is not utilized in the training of the RL agent. Therefore, we propose two enhancements that improve the sample efficiency of the SAC-RCBF framework: (i) utilizing a differentiable version of the safety layer which allows for backpropagation through the QP and thus to explicitly account for the QP's output in the RL losses, and (ii) leveraging the partially learned dynamics to generate synthetic data that can be added to the replay buffer. It is important to note that, although we assume knowledge of the output dynamics of the system necessary for safety, it may be the case that the full dynamics or the reward function are unknown and that the safety considerations are only known locally (e.g., obstacle detection). In such scenarios, the generation of synthetic data may be infeasible and only the benefits of the differentiable safety layer can be leveraged. In what follows, we discuss both of these approaches and the final framework is presented in Algorithm~\ref{alg:2}. 

\subsubsection{Differentiable Safety Layer}

As discussed in the Introduction, the final actions taken in the environment are the altered safe actions obtained from the RCBF layer. However, the updates in equations \eqref{eq:q_update} and \eqref{eq:pi_update} are typically taken with respect to the potentially unsafe RL actions. This is equivalent to the effects of the safety layer being part of the transition dynamics with respect to the RL algorithm. As such, large changes in the output of the safety layer, which typically occur near unsafe states, can render learning difficult. Therefore, explicitly accounting for the safety layer's output in the RL losses can significantly improve the learning performance.

To achieve this, we propose utilizing the differentiable optimization framework introduced in \cite{amos2017optnet, agrawal2019differentiable}, which enables the propagation of gradients through the QP as in \eqref{eq:QP} using a linear system formulated from the KKT conditions. As such, we can take the updates in \eqref{eq:q_update} and \eqref{eq:pi_update} explicitly with respect to the safe action and backpropagate through the RCBF-QP in an end-to-end fashion which improves the learning performance as discussed in Section~\ref{sec:experiments}. The resulting losses for SAC are thus given by
\begin{align}
\label{eq:q_update_diff}
    J_{Q}(\theta) &= \mathbb{E}_{(x_t, u^{RL}_t, u^{S}_t) \sim \mc{D}_R} [ \frac{1}{2}  (Q_{\theta}(x_t, u^{RL}_t + u^{S}_t) \\
    &- (r(x_t, u^{RL}_t + u^{S}_t) + \gamma \mathbb{E}_{x_{t+1} \sim p} [V_{\bar{\theta}}(x_{t+1})]))^{2}],
\end{align}
where
\begin{equation}
V_{\bar{\theta}}(x_{t}) = \mathbb{E}_{ u^{RL}_t \sim \pi_{\phi}} [ Q_{\bar{\theta}}(x_t, u^{RL}_t + u^{S}_t) - \alpha_{e} \log{\pi_{\phi}(u^{RL}_t | x_t)}], 
\end{equation}
and
\begin{align}
\label{eq:pi_update_diff}
J_{\pi}(\phi) &= \mathbb{E}_{ x_t \sim \mc{D}_R} [\mathbb{E}_{ u^{RL}_t \sim \pi_{\phi}} [\alpha_{e} \log{\pi_{\phi}(u^{RL}_t | x_t)} \\
&\quad- Q_{\theta}(x_t, u^{RL}_t + u^{S}_t)]].
\end{align}
Note that the explicit dependencies of each variable have been removed for brevity. Moreover, although directly accounting for the effects of the RCBF during training results in increased computation time because batches of QPs are being solved during the computation of the losses \eqref{eq:q_update_diff} and \eqref{eq:pi_update_diff}, the inference time remains the same since only a single QP is solved at each time step.

\subsubsection{Model-Based RL}
\begin{algorithm}[t]
	\caption{SAC-RCBF}
	\label{alg:2}
 	\begin{algorithmic}[1]
	\Require
		\Statex Dynamics prior $f(\cdot)$ and $g(\cdot)$ and RCBF $h(\cdot)$ 
	\For{$N$ iterations}
		\State Train GP models $p_{\psi}$ on $\mathcal{D}_{\text{env}}$  
		\For{$E$ environment steps}
		    \State Obtain action $u^{\text{RL}}_t$ from $\pi_{\phi}$
	        \State Obtain $u^{S}_t$ using $h$ and $p_{\psi}$ \Comment \eqref{eq:QP}
            \State Take action $u^{\ast}_t$ in environment \Comment \eqref{eq:final_action}
            \State Add transition $(x_t, u^{\ast}_t, x_{t+1}, r_t)$ to $\mc{D}_{\text{env}}$
        \EndFor
        \If {Model-Based}
		\For {$M$ model rollouts}
		    \State Sample $x_t$ uniformly from $\mc{D}_{\text{env}}$
		    \For {$k$ model steps}
		        \State Obtain action $u^{\text{RL}}_t$ from $\pi_{\phi}$
		        \State Obtain $u^{S}_t$ using $h$ and $p_{\psi}$ \Comment \eqref{eq:QP}
		        \State Get synthetic transition using $u^{\ast}_t$ and $p_{\psi}$
		        \State Add transition $(x_t, u^{\ast}_t, x_{t+1}, r_t)$ to $\mc{D}_{\text{model}}$
		    \EndFor
		\EndFor
		\EndIf
		\For{$G$ gradient steps}
		   \If {Model-Based} 
		    \State Sample batch $B$ from $\mc{D}_{\text{model}}$
		   \Else
		   \State Sample batch $B$ from $\mc{D}_{\text{env}}$
		   \EndIf
		  \State Update ($\phi$ and $\theta$) using $B$ \Comment \eqref{eq:q_update}, \eqref{eq:pi_update}
		\EndFor
	\EndFor
 	\end{algorithmic}
\end{algorithm}
When possible, i.e. in cases where the reward function and safety considerations are known ahead of time, to further increase the sample efficiency of SAC-RCBF, we leverage the partially learned dynamics, the reward function and the RCBF constraints to generate short-horizon rollouts as in \cite{janner2019trust}. The use of short-horizon rollouts is motivated by the fact that model errors compound over long horizons, and thus the data collected from such rollouts would not benefit the learning and, in the worst case, can significantly impede the agent from learning an optimal policy. 

Specifically, as highlighted in Algorithm~\ref{alg:2}, at each iteration, we sample $M$ initial states from transitions in the environment's replay buffer $\mathcal{D}_{\text{env}}$ and generate new, synthetic $k$-step rollouts using the dynamics prior and the disturbance estimates from the GPs. In turn, these newly generated transitions are added to another replay buffer $\mathcal{D}_{\text{model}}$ which is used to train the agent.

It is important to note that, in cases where the reward function and safety considerations are known ahead of time it is also reasonable to leverage Model Predictive Control (MPC) instead of RL. However, as opposed to MPC which is typically only amenable for real-time control-synthesis if the resulting optimization is convex, using RL has the advantage of being computationally efficient at test time once the policy is learned. Moreover, leveraging RL also has the additional benefit that we obtain a value function at the end of training, which we can be later leveraged as a Control Lyapunov Function (CLF) or CBF itself \cite{ohnishi2021constraint}.  

\subsection{Modular Task Learning}
\label{subsec:mod-task-learning}
When deploying RL on real robots, certain applications may require different constraints during the training stage as compared to testing. For example, consider the case where a differential-drive robot is tasked with navigating to a goal location subject to avoiding obstacles. It may be the case that locations and types of the obstacles differ between the training environment and the potential test environments. However, as we will demonstrate in the experiments, the safety considerations during training directly affect both the learned value function and the policy which is undesirable for transfer. In the differential-drive robot's case, this would correspond to undesired low values assigned around the obstacles encountered during training. This is due to the modification of the RL policy's action by the safety layer, which directly affects the state transition. Thus, in this section, we present a simple yet effective approach, which we refer to as Modular SAC-RCBF, for learning the value function and optimal policy of the reward-driven task modularly, i.e., independently of a subset of the constraints during training.

The effect of the safety layer can be directly observed in the Bellman update equation
\begin{equation}
\label{eq:safe-bellman}
\begin{aligned}
Q(x_t, u^{RL}_t + u^{S}_t) &=  r(x_t, u^{RL}_t+u^{S}_t) \\&+ \gamma \mathbb{E}_{x_{t+1} \sim p}[ Q(x_{t+1}, u^{RL}_{t+1} + u^{S}_{t+1})].
\end{aligned}
\end{equation}
 Specifically, note that $x_{t+1}$ is a direct function of $u^{S}_t$, and that $u^{S}_{t+1}$ affects the final action taken at $x_{t+1}$ when considering the cost to go. Due to the former, entirely discarding the effects of the modifications to the action by the safety layer (i.e., setting all $u^{S}=0$ in \eqref{eq:safe-bellman}) in the loss functions \eqref{eq:q_update_diff} and \eqref{eq:pi_update_diff} does not yield the desired Q-function since the cost to go is still dependent on $x_{t+1}$ through the dynamics \eqref{eq:discrete_dynamics}. 

Thus, the modification we propose is to solely discard the safety's layer output when computing the cost-to-go (i.e., only setting $u^{S}_{t+1}=0$ in \eqref{eq:safe-bellman}). The resulting loss for the Q-function is given by
\begin{align}
\label{eq:q_update_mod}
    J_{Q}(\theta) &= \mathbb{E}_{(x_t, u^{RL}_t, u^{S}_t) \sim \mc{D}_R} [ \frac{1}{2}  (Q_{\theta}(x_t, u^{RL}_t + u^{S}_t) \\
    &- (r(x_t, u^{RL}_t + u^{S}_t) + \gamma \mathbb{E}_{x_{t+1} \sim p} [V^{mod}_{\bar{\theta}}(x_{t+1})]))^{2}],
\end{align}
where
\begin{equation}
V^{mod}_{\bar{\theta}}(x_{t}) = \mathbb{E}_{ u^{RL}_t \sim \pi_{\phi}} [ Q_{\bar{\theta}}(x_t, u^{RL}_t) - \alpha_{e} \log{\pi_{\phi}(u^{RL}_t | x_t)}], 
\end{equation}
and the resulting policy loss is identical to \eqref{eq:pi_update} as it no longer accounts for the effects of the safety layer.

Intuitively, \eqref{eq:q_update_mod} ignores the constraint considerations when computing the cost-to-go during training. For instance, consider the example mentioned above, where a differential-drive robot is to navigate safely through obstacles to reach a desired location. This approach would allow the robot to consider the cost-to-go as if there were no obstacles and, consequently, to learn reaching the goal (i.e., the reward-driven task) in a modular fashion. It is important to note that this method does not affect the safety of the agent during training since the final action taken in the environment given by \eqref{eq:final_action} remains unchanged. Moreover, we demonstrate empirically in the next section that this approach allows the RL agent to learn a value function that is independent of the safety layer, and, consequently, significantly improves performance at zero-shot transfer tasks.   

\section{Experiments}
\label{sec:experiments}

In this section, we present two sets of experiments that validate the efficacy of the proposed methods \footnote{The code can be found at \href{https://github.com/yemam3/Mod-RL-RCBF}{github.com/yemam3/Mod-RL-RCBF}.}. The first set of experiments aims to demonstrate the improvement in sample efficiency gained by leveraging a differentiable safety-layer as well as utilizing the partially learned model to generate synthetic data. On the other hand, the second set of experiments serves to validate the method proposed in Section~\ref{subsec:mod-task-learning} for modularly learning the reward-driven task. 
\vspace{-1mm}
\subsection{Sample Efficiency}
In this subsection, we compare the sample efficiency of both the Model-Based (MB) and Model-Free (MF) versions of the proposed SAC-RCBF approaches against two baselines in two custom environments. Specifically, the first baseline only utilizes model-free SAC and a non-differentiable RCBF-QP safety layer, whereas the second baseline also leverages the compensator from \cite{cheng2019end} which is a neural-net trained to learn the behavior of the RCBF-QP during training. We note that the reason we did not utilize discrete-CBFs as in \cite{cheng2019end} for the second baseline is that they are only amenable for real-time control-synthesis for affine functions which is restrictive as discussed in the Introduction. In both environments, the metric used for comparison is the agents' performances (i.e. the cumulative sum of rewards over an episode) with respect to the number of episodes. Moreover, we note that no safety violations occurred throughout the experiments, thus validating the effectiveness of the RCBF-QP at keeping the system safe throughout training. 

\subsubsection{Unicycle Environment}
\label{subsec:unicycle_env}

The first environment, inspired by the Safety Gym environment \cite{ray2019benchmarking}, consists of a unicycle robot tasked with reaching a goal location while avoiding obstacles. We chose to build a custom environment due to the fact that explicit knowledge of $f$ and $g$ from \eqref{eq:main_system} is needed for the formulation of the RCBF constraint, which is not the case for Safety Gym.  Note that, in this section, the explicit dependence on time is dropped for brevity.

The robot is modelled using disturbed unicycle dynamics as 
\begin{equation}
    \dot{x} = 
    \begin{bmatrix}
    \dot{x_1} \\ \dot{x_2} \\ \dot{\theta} 
    \end{bmatrix}
    =
    \begin{bmatrix}
    \cos{\theta} & 0 \\ 
    \sin{\theta} & 0 \\
    0 & 1 
    \end{bmatrix} 
    (u + u_d), 
\end{equation}
where $u =  
\begin{bmatrix}
    v & \omega
\end{bmatrix}^{\top}$
is the control input with $v$ and $\omega$ denoting the linear and angular velocities of the robot respectively, and $u_d = -0.1\cos(\theta) \begin{bmatrix} 1 ~ 0 \end{bmatrix}^{\top}$ is an unknown disturbance aimed at simulating a sloped surface. In turn, from the agent's perspective, the dynamics are modelled through the following differential inclusion
\begin{equation}
   \dot{x} \in
    \begin{bmatrix}
    \cos{\theta} & 0 \\ 
    \sin{\theta} & 0 \\
    0 & 1 
    \end{bmatrix} 
    u + D(x), 
\end{equation}
where the disturbance set $D(x)$ is learned via GPs.
\begin{figure}
    \centering
    \includegraphics[width=0.58\linewidth]{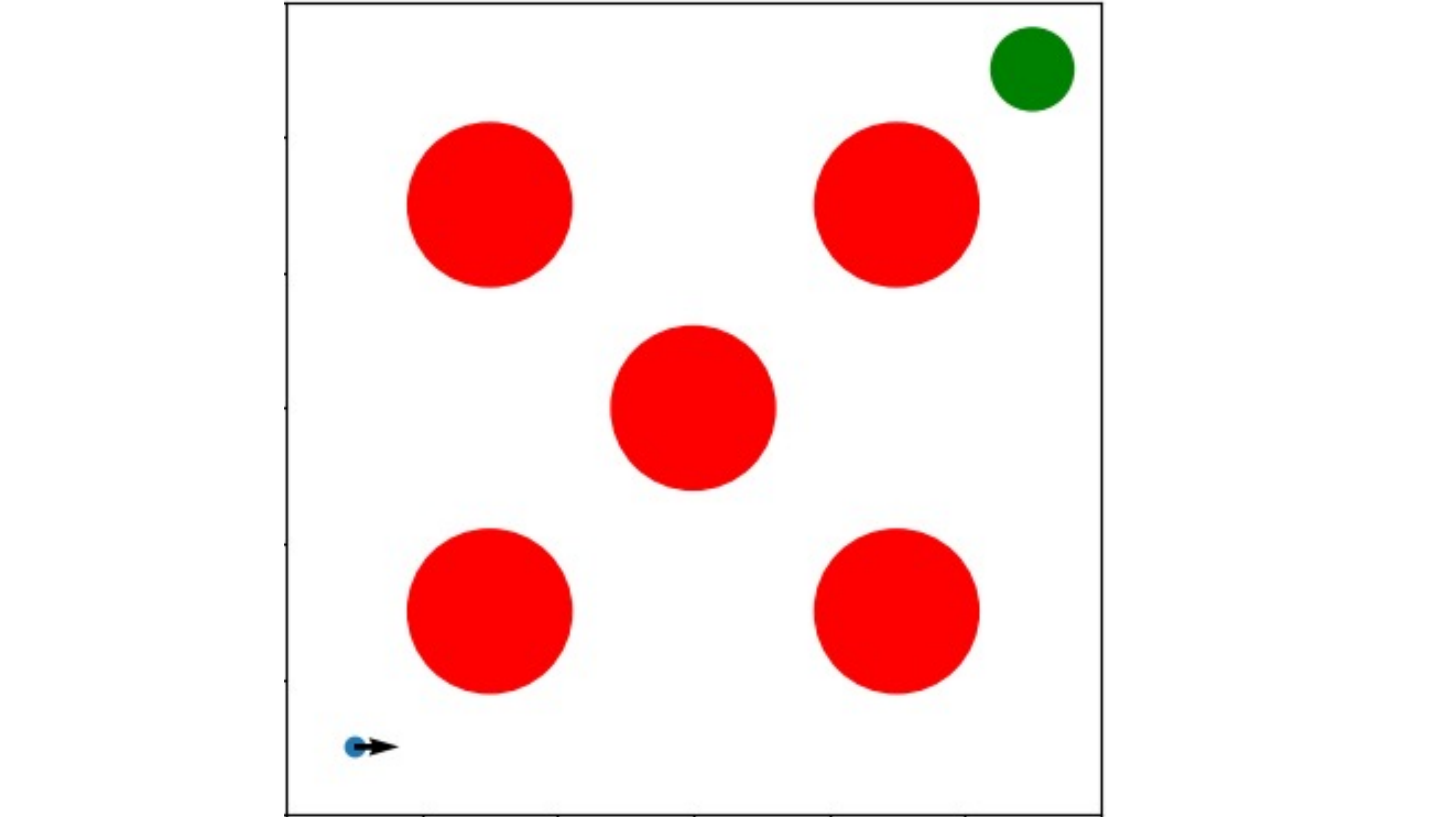}
    \caption{Snapshot of the Unicycle environment. The agent (blue) is tasked with reaching a desired location (green) while avoiding the obstacles (red).}
    \label{fig:safety-gym}
\end{figure}
\begin{figure}
    \centering
    \includegraphics[width=0.80\linewidth]{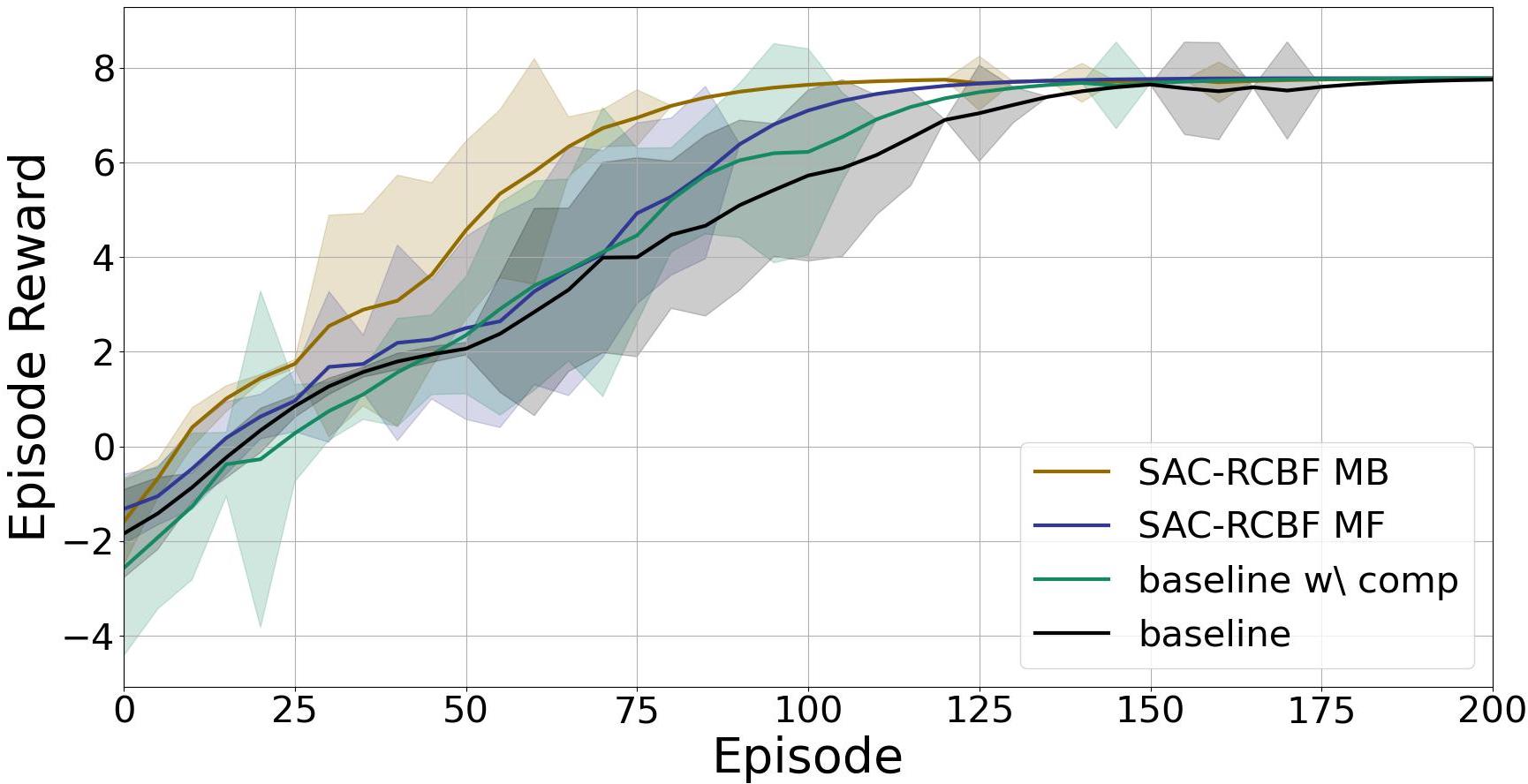}
    \caption{Comparison of episode reward versus number of training episodes in the Unicycle environment for the Model-Based (MB) and Model-Free (MF) versions of SAC-RCBF and the two baselines. Each plot is an average of ten experiments using different seeds, and the light shadowing denotes a confidence interval of two standard deviations.}
    \label{fig:unicycle_env_rew}
\end{figure}
Similar to \cite{emam2019robust}, we can encode a collision-avoidance constraint between the agent and a given obstacle $o$ using the following RCBF $h_{o}(p(x)) = \norm{p(x) - p_{o})}^2 - \delta^{2}$ ,
where $\delta > 0$ denotes the minimum desired distance between the center of the robot $p(x)$ and the position of the obstacle $p_{o}$.

Shown in Figure~\ref{fig:unicycle_env_rew} is a plot of the episode rewards versus training episodes for the various approaches. As shown, all approaches converge to an optimal policy and MB SAC-RCBF performs best in terms of sample efficiency which highlights the benefit of generating synthetic data. On the other hand, the MF SAC-RCBF approach performs only slightly better than the two baselines which indicates that solely leveraging the differentiable safety layer has little benefit in this easier environment. The latter can be attributed to the fact that the behavior of the safety layer is smooth in this case as the obstacles are circular and thus the baselines, which do not explicitly account for the RCBF-QP's behavior, are able to learn the optimal policy with comparable sample efficiency.     

\subsubsection{Car Following}

The second environment is based on \cite{cheng2019end} and involves a chain of five cars driving in a lane each modelled as a disturbed double integrator
\begin{equation}
    \dot{x}_i = \begin{bmatrix}
        \dot{p}_i \\
        \dot{v}_i
    \end{bmatrix}
    = \begin{bmatrix}
        0 & 1 \\
        0 & 0 \\
    \end{bmatrix} x_i + 
    \begin{bmatrix}
        0 \\
        1 + d_i
    \end{bmatrix} u_i,
\end{equation}
where $p_i, v_i, u_i$ are the position, velocity and input acceleration of car $i$ respectively. In addition, $d_i = 0.1$ is a disturbance unknown to the agent that is non-zero for all cars except car $4$ (i.e., $\forall i \in \{1,2,3,5\}$).

The agent only controls the acceleration of the fourth car and can observe the positions, velocities and accelerations of all five cars. The leading car's behavior is aimed at simulating traffic through a sinusoidal acceleration $u_1 = 10 \sin(0.2t)$. Cars $2$ and $3$ behave as follows
\begin{align*}
    u_i = 
    \begin{cases} 
      k_v (v_{\text{des}} - v_i) - k_b (p_{i-1} - p_i)  & \text{if} \; |p_{i-1} - p_i| < 6  \\
      k_v (v_{\text{des}} - v_i)  & \text{o.w.},
   \end{cases}
\end{align*}
where $v_{des} = 30$ is the desired velocity and $k_v$ and $k_b$ are the velocity and braking gains respectively. Lastly, the last car in the chain (i.e., car $5$) has the following control input
\begin{align*}
    u_5 = 
    \begin{cases} 
      k_v (v_{\text{des}} - v_5) - k_b (p_{3} - p_5)  & \text{if} \; |p_3 - p_5| < 12  \\
      k_v (v_{\text{des}} - v_5)  & \text{o.w.}.
   \end{cases}
\end{align*}
As such, the objective of the RL agent is to minimize the total control effort, which is captured through the reward function $r(u)~=~-||u||^2$, while avoiding collisions. Moreover, two RCBFs are formulated which enforce a minimum distance between the cars three and four, and four and five respectively
$
    h_{i}(x) = \frac{1}{2} (|p_i - p_4|^2 - \delta^2), \;\forall i \in \{3,5\},
$
where $\delta = 3.5$ is the minimum distance. We note that since $h_{3}$ and $h_{5}$ have relative degree $2$, a cascaded RCBF formulation as in \cite{notomista2020persistification} is leveraged to obtain the collision avoidance constraints. Specifically, we define 
$
    h^{\ast}_{i}(x) = \dot{h}_i(x) + \gamma_1 h_i(x), \; \forall i \in \{3,5\},
$ and enforce the forward invariance of the super-zero level sets of $h^{\ast}_{i}$ through RCBFs using a similar procedure to the one described above.  The theoretical particulars of cascaded RCBFs will be discussed in future work, and we present the results without proof here.
\begin{figure}
    \centering
    \includegraphics[width=0.80\linewidth]{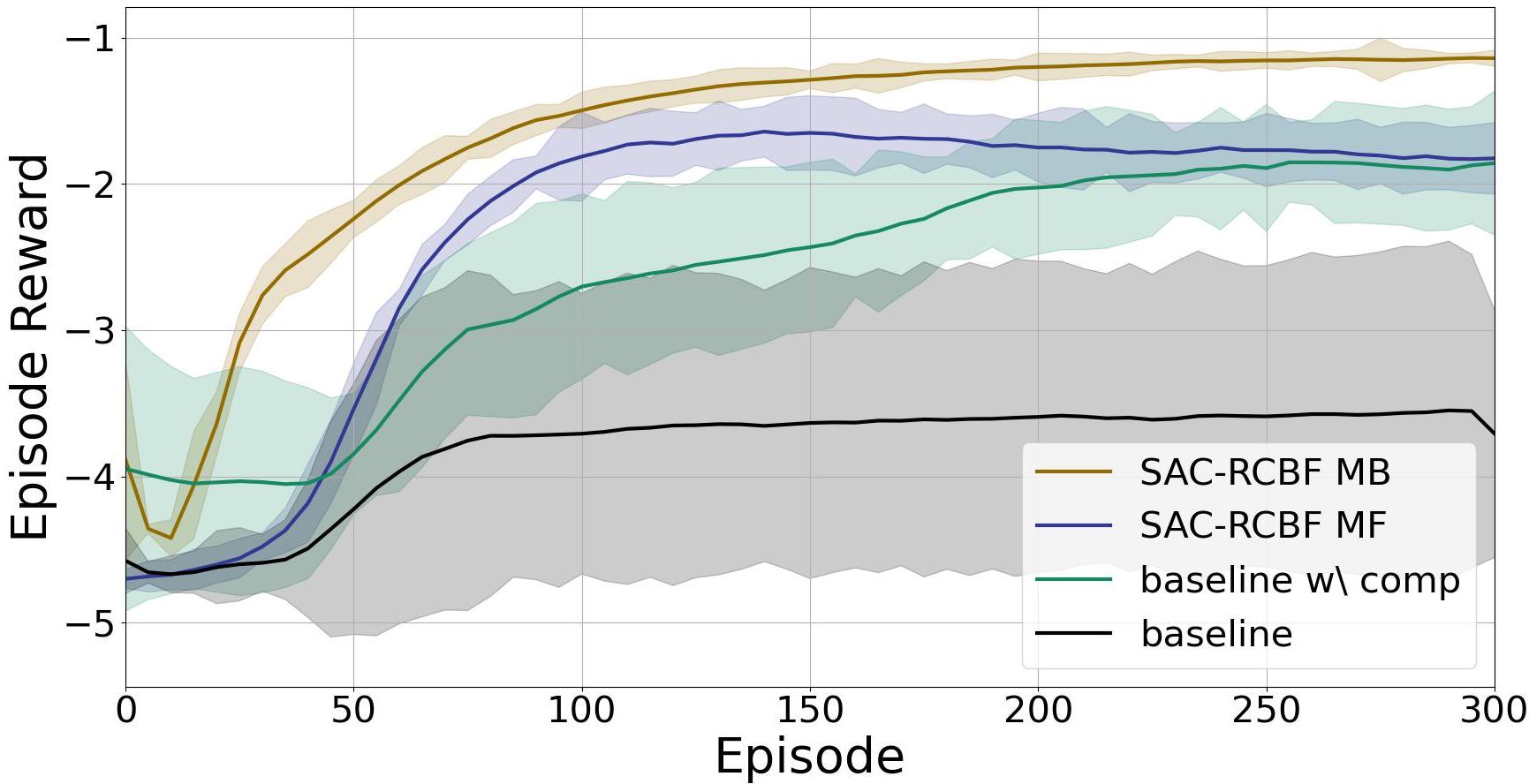}
    \caption{Comparison of episode reward versus number of training episodes in the Car Following environment for the Model-Based (MB) and Model-Free (MF) versions of SAC-RCBF and the two baselines. Each plot is an average of ten experiments using different seeds, and the light shadowing denotes a confidence interval of two standard deviations.}
    \label{fig:car_env_rew}
\end{figure}
Shown in Figure~\ref{fig:car_env_rew} is a plot of the episode rewards versus training episodes for the different approaches. As opposed to the previous environment, both the MB and MF SAC-RCBF frameworks exhibit significantly better sample efficiency compared to the baselines. This can be attributed to the fact that the behavior of the RCBF-QP in this environment strongly affects both the dynamics and the rewards collected at each episode. Specifically, the RCBF-QP layer can cause the controlled car (Car 4) to decelerate or accelerate rapidly. Such large change in the QP's output, if not accounted for explicitly, is equivalent to a sharp transition in the dynamics which is difficult to learn. This is further highlighted by the inability of the baseline without the compensator at learning a meaningful policy in several trials as shown in Figure~\ref{fig:car_env_rew}. In turn, this signifies that explicitly accounting for the safety layer's behavior is not only critical for sample-efficient training but also for finding an optimal policy.
\vspace{-2.0mm}
\subsection{Modular Task Learning}
In this set of experiments, we aim to validate the Modular SAC-RCBF approach proposed in Section~\ref{subsec:mod-task-learning} by visualizing the learned value functions for various approaches and comparing the zero-shot transfer performance in two different environments. 

\subsubsection{Unicycle}

\begin{table}[tb]
    \centering
    \small
    \caption{Zero-shot transfer to $200$ different obstacle configurations in the Unicycle environment.}
    \resizebox{\columnwidth}{!}{%
    \begin{tabular}{|c|c|c|c|}
        \hline
        Approach  & Mean Ep. Rew. & Std Ep. Rew. & Mean Comp. \\
        \hline
        SAC w/o obstacles (upper bound) & $7.009$ & $1.745$ & $0.81$ \\ 
        Baseline & $3.533$ & $2.685$ & $0.21$  \\ 
        SAC-RCBF & $3.223$ & $2.413$ & $0.17$  \\ %
        Mod. SAC-RCBF & $6.189$ & $2.317$ & $0.63$ \\ 
       \hline
    \end{tabular}
    }
    \label{tab:0-shot-unicycle}
\end{table}

\begin{figure}
    \centering
    \includegraphics[width=0.99\linewidth]{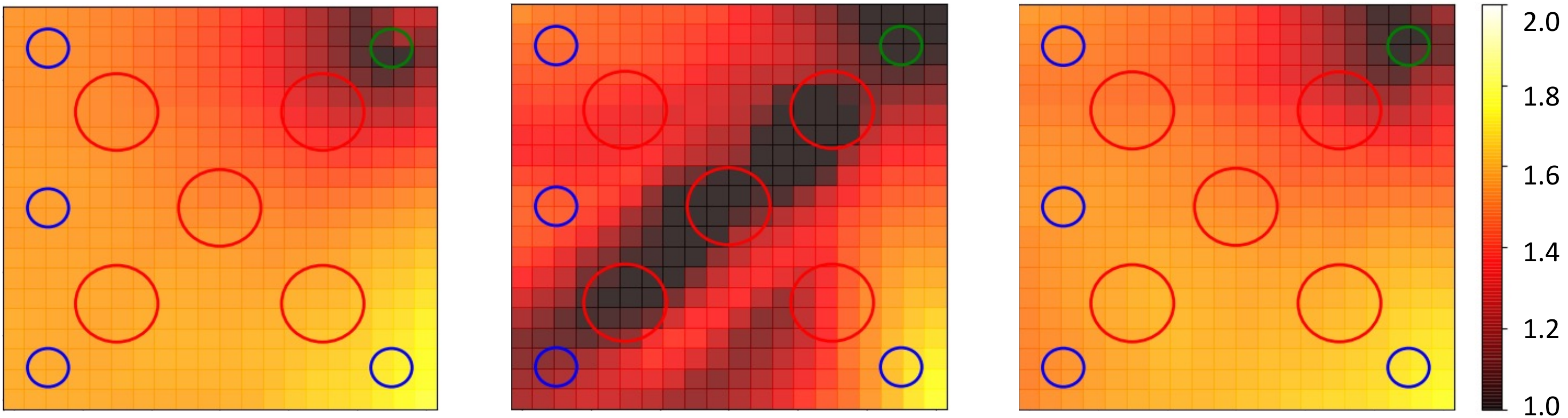}
    \caption{Discretized plot of the learned value function for SAC trained with no obstacles and no safety layer (left), SAC-RCBF (middle), Modular SAC-RCBF (right). The blue, red and green circles denote the possible starting locations for each episode, the obstacles, and the goal respectively. The Modular SAC-RCBF approach learns a value function that is almost identical to SAC trained with no obstacles, whereas the SAC-RCBF approach differs in that it assigns low values around the obstacles.}
    \label{fig:unicycle-v-plots}
\end{figure}

In the Unicycle environment, the objective is to learn a modular value function for the reward-driven task (i.e., go-to-goal) that is independent of the obstacle configuration present during training. Therefore, to visually validate the efficacy of Modular SAC-RCBF, shown in Figure~\ref{fig:unicycle-v-plots} are the learned value functions for SAC-RCBF (middle) and Modular SAC-RCBF (right) trained with the obstacles denoted in red. By visually comparing these learned value functions to the target value function, i.e., one learned by SAC trained in an environment with no obstacles and no safety layer (left), it is clear that SAC-RCBF assigns undesirable low values around the obstacles. We note that the baselines from the previous set of experiments also assign undesirable low values around the obstacles and their value function visualizations have been omitted for brevity. On the other hand, Modular SAC-RCBF learns a value function that is almost identical to the target making it suitable for transfer.

To further validate the approach, we tested the zero-shot transfer ability of the learned policies in $200$ environments with different obstacle shapes, sizes and locations and thus different constraints in the safety layer. Shown in Table~\ref{tab:0-shot-unicycle} are the mean and standard deviation of episode rewards as well as the proportion of completed (i.e., if the robot reached the goal) environments for the various approaches. As expected SAC trained with no obstacles and no safety layer solves the most environments and had a completion rate of $81\%$ which can be treated as an upper-bound, followed by Modular SAC-RCBF which had a completion average of $63\%$. On the other hand, both the baseline and SAC-RCBF performed quite poorly solving only $21\%$ and $17\%$ of the environments respectively. These performances were mainly due to the effect that the obstacles during training had on the learned value functions and, consequently, the optimal policies.   

\subsubsection{PVTOL}
In this experiment, we test the zero-shot transfer ability of the various approaches in an environment involving a 2D quadcopter tasked with navigating to a goal location while avoiding obstacles and staying within a certain distance of a safety pilot during training. 

We utilize a modified PVTOL model given by
\begin{equation*}
[
    \dot{x_1} \:
    \dot{x_2} \:
    \dot{v_1} \:
    \dot{v_2} \:
    \dot{\theta} \:
    \dot{f} 
]^\top = 
[    v_1 \;
    v_2 \;
    -\sin(\theta)f \;
    \cos(\theta)f - 1 \;
    \omega + d_{\omega} \;
    u_f
]^\top,
\end{equation*}
where $x_i$, $v_i$ denote the position and velocity along axis $i$ respectively, $\theta$ is the orientation of the robot and $f$ is the thrust. Differently than the traditional PVTOL model \cite{zavala2003global}, we assume direct control of the angular velocity $\omega$ and the derivative of the thrust denoted by $u_f$ so as to permit the formulation of the safety RCBFs. We note that these assumptions are reasonable since a popular controller design approach for PVTOLs is to use the moment $M$ to track a desired $\omega$, and we model the tracking error as a disturbance on $\omega$'s dynamics denoted by $d_{\omega}$. Moreover, controlling the derivative of the thrust can be straightforwardly achieved by augmenting the system with artificial thrust dynamics as performed here. 

As such, the RCBF utilized for obstacle avoidance is given by $h_{o}(p(x)) = \norm{p(x) - p_{o})}^2 - \delta^{2}$,
where $\delta > 0$ denotes the minimum desired distance between $p(x)~=~[x_1 x_2]^{\top}$ and the position of the obstacle $p_{o}$. Similar to the Car Following environment, since $h_{o}$ has relative degree $3$, a cascaded RCBF formulation as in \cite{notomista2020persistification} is leveraged to obtain the collision avoidance constraints. The safety operator is assumed to have noisy single-integrator dynamics along the $x_1$-axis and tracks the drone via a low-gain proportional controller. The RCBF enforcing the 2D quadcopter to keep a certain distance from the operator is given by $h_{h}(x_1)~=~\delta^{2} - (x_1 - x_h)^2$.

\begin{table}[tb]
    \centering
    \caption{Zero-shot transfer to $200$ different obstacle configurations in the PVTOL environment.}
    \resizebox{\columnwidth}{!}{%
    \begin{tabular}{|c|c|c|c|}
        \hline
        Approach  & Mean Ep. Rew. & Std Ep. Rew. & Mean Comp. \\
        \hline
        SAC w/o operator/obs. (upper bound) & $11.969$ & $2.376$ & $0.87$ \\ 
        Baseline & $7.228$ & $4.622$ & $0.34$  \\
        SAC-RCBF & $9.450$ & $3.808$ & $0.52$  \\
        Mod. SAC-RCBF & $10.656$ & $3.411$ & $0.69$ \\
       \hline
    \end{tabular}
    }
    \label{tab:0-shot-pvtol}
\end{table}

Shown in Table~\ref{tab:0-shot-pvtol} are the zero-shot transfer results of the policies learned by the different methods to $200$ randomized environments with different numbers and locations of obstacles as well as the presence of a safety operator constraint. Similar to the Unicycle environment, we obtain a performance upper-bound by training SAC in an environment with no obstacles nor a safety operator and thus without a safety layer. As expected, this approach  performed best at zero-shot transfer with a completion rate of $87\%$. Moreover, Modular SAC-RCBF yields a $33\%$ improvement over SAC-RCBF by completing $69\%$ of the environments as compared to $52\%$. As such, this experiment validates that Modular SAC-RCBF is suitable for transfer tasks over CBF constraints.  

\section{Conclusion}
\label{sec:conclusion}
In this paper, we introduce a novel sample-efficient framework that combines RL with a RCBF-based layer to enforce safety during training. The resulting SAC-RCBF approach is empirically validated in simulated environments and is shown to improve both the sample efficiency and steady-state performance during RL training. Moreover, we propose an approach for modularly learning the reward-driven task independently of a subset of the safety constraints resulting in learned policies that can be leveraged for transfer to different environments involving different safety constraints. A relevant direction of future work is to utilize the proposed framework for the reinforcement learning of non-safety critical CBFs.





\bibliographystyle{IEEEtran}
\bibliography{ref}


 





\end{document}